\documentclass[preprint,12pt]{elsarticle}

\usepackage{graphicx}	 		
\usepackage{amsmath,amsfonts,amsbsy}
\usepackage{slashed}

\usepackage[scr=boondoxo,scrscaled=1.05]{mathalfa}

\usepackage{hyperref}

\newcommand{\laser}{\mathcal{A}}
\newcommand{\laserforslash}{\mathcal{A}\,\,}
\newcommand{\psl}{\slashed{p}}
\newcommand{\ksl}{\slashed{k}}
\newcommand{\lasersl}{\slashed{\laserforslash}\!\!} 
\newcommand{\ssl}{\slashed{s}}
\newcommand{\lasers}{\laser^{*}}
\newcommand{\laserssl}{\slashed{\laser^*}}

\newcommand{\intfsDuv}{\int_{_\mathrm{UV}}{\bar{}\kern-0.45em d}^{\,D\!}s\,}

\newcommand{\Ab}{\mathrm{A}}
\newcommand{\Em}{\mathrm{E}}
\newcommand{\Prop}[1]{\mathrm{P}_{#1}}
\newcommand{\In}{\mathrm{I}}
\newcommand{\Out}{\mathrm{O}}
\newcommand{\duv}{\delta_{_{^\mathrm{UV}}}}
\newcommand{\pk}{p{\cdot}k}

\newcommand{\plaser}{p{\cdot}\laser}
\newcommand{\plasers}{p{\cdot}\lasers}

\newcommand{\laserm}{\lasers\!{\cdot}\laser }

\newcommand{\mstar}{\mathscr{M}}
\newcommand{\mstarsl}{\slashed{\mstar}}
\newcommand{\sigmasl}{\slashed{\Sigma}_{_{\!\!\mstar}}}

\begin{document}

\begin{frontmatter}

\title{One loop Volkov propagator in the Lorentz class of gauges}

\author{Martin~Lavelle and David~McMullan}

\address{Centre for Mathematical Sciences\\University of Plymouth \\
Plymouth, PL4 8AA, UK\\
mlavelle@plymouth.ac.uk\qquad dmcmullan@plymouth.ac.uk}

\begin{abstract}
We calculate the ultraviolet divergences in a weak field expansion of the Volkov propagator. This is done for the full Lorentz class of gauges. The expected gauge invariance of the vacuum mass shift in each sideband is recovered. However, the renormalisation of the background induced mass shift is shown to be gauge dependent. In particular, we show that it vanishes in Landau gauge. We find that only in that gauge does the vacuum renormalisation remove all ultraviolet divergences. 
\end{abstract}


\end{frontmatter}

%\maketitle
The Volkov solution~\cite{Volkov:1935zz}  of an electron in a plane wave background is the paradigm for models of charged matter in a laser background.   One of the most striking features of this model is the laser induced mass shift at tree level. Additionally, the electron two point function exhibits a sideband structure consisting of a sum over propagators where the momentum is shifted by integer multiples of the laser momentum~\cite{Reiss:1966A}. This solution has been developed for a wide class of polarisations~\cite{Lavelle:2017dzx}. We note that these results are all obtained by using  the background field gauge condition  $k\cdot\laser=0$, where $\laser_\mu$ is the background potential and $k^\mu$ is the null momentum pointing along the laser direction. 
 
Loop corrections in a laser background have been looked at several times, as for example in~\cite{Becker:1974en}\cite{Baier:1975ys}\cite{baier75a}\cite{Brouder:2002fz}\cite{Milstein:2006zz}\cite{ Heinzl:2011ur}\cite{DiPiazza:2011tq}\cite{Meuren:2011hv}\cite{Dinu:2013gaa}\cite{Meuren:2013oya}\cite{King:2015tba}\cite{Gies:2016yaa}\cite{King:2018wtn}\cite{Podszus:2018hnz}\cite{Ilderton:2019kqp}.  
In a recent paper~\cite{Lavelle:2019lys}, we have calculated the ultraviolet divergences for the electron two point function to one loop in a weak field expansion of the Volkov solution. Among other things, we showed there that, as well as the usual vacuum renormalisation of the wave function and vacuum mass shift, an additional renormalisation of the background induced mass shift was required. The loop calculations in that paper, as is commonplace in the literature, were performed in Feynman gauge. The aim of this paper is to extend those calculations to the full Lorentz (or $R_\xi$) class  of gauges. See also \cite{DiPiazza:2018ofz} for a discussion of gauge transformations and the Volkov propagator.  
 
As discussed in, for example, Section 8.5 of \cite{Schwartz:2013pla},  the  photon propagator in the  Lorentz class of gauges is  given by
\begin{equation}
  	D_{\mu\nu}(s)=\frac{-i}{s^2+i\epsilon}\Big(g_{\mu\nu}+(\xi-1)\frac{s_\mu s_\nu}{s^2}\Big)\,,
  \end{equation}
where Feynman gauge corresponds to $\xi=1$ and Landau (or Lorenz) gauge to $\xi=0$. In a laser background, the tree level electron propagator of momentum $p$ picks up multiples of the laser momentum, $k$. We write this propagator after $n$ net absorptions as
\begin{equation}
	\Prop{n}=\frac{i}{\psl+n\ksl-m+i\epsilon}\,.
\end{equation}
Loop corrections to this will introduce a gauge dependence that we will calculate below. 

\begin{figure}[htb] 
\[
	\includegraphics{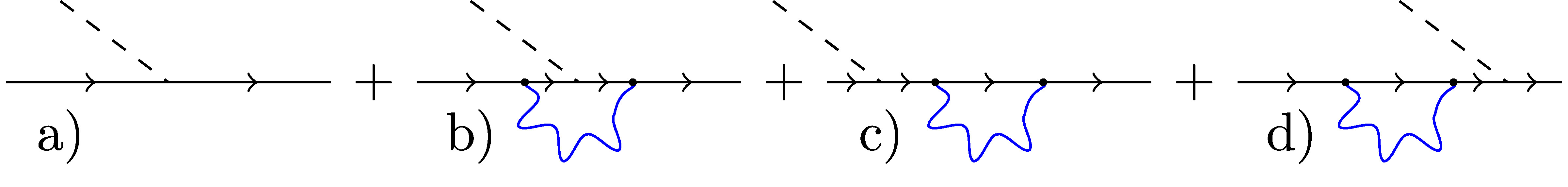} 
\]

\caption{Laser absorption by the electron and its one-loop corrections. }
	\label{fig:in_plus_loop}
\end{figure}

The absorption of a laser photon by the electron is given by the vertex
\begin{equation}\label{eq:Ais}
	\Ab=-i\lasersl\,.%\,,\qquad\Em=-ie\laserssl\,,
\end{equation}
Note that a factor of the coupling has been absorbed into the definition of the background field, see the discussion in~\cite{Lavelle:2019lys} for more details. The tree level absorption of Fig.~\ref{fig:in_plus_loop}~a) is then given by $\Prop{n+1}\Ab\Prop{n}$. This can be rewritten in terms of the difference of two propagators by using the  absorption identity 
\begin{equation}\label{eq:AbsId}
	\Prop{n+1}\Ab\Prop{n}=\In\Prop{n}-\Prop{n+1}\In\,,
\end{equation}  
where the \lq In\rq\ factor is given by the matrix term
\begin{equation}
	\In=\frac{2\plaser+\ksl\lasersl}{2\pk}\,.
\end{equation}
This ability to write an interaction with the background in terms of a sum of propagators is generically called a sideband description.

The dual emission process is characterised by the  vertex $\Em=-i\laserssl$. This leads to the corresponding dual emission identity $\Prop{n}\Em\Prop{n+1}=\Prop{n}\Out-\Out\Prop{n+1}$, where the \lq Out\rq\ factor is
\begin{equation}
	\Out=\frac{2\plasers+\laserssl\ksl}{2\pk}\,.
\end{equation}

The loop corrections to the absorption process in Fig.~\ref{fig:in_plus_loop} are of two types: a vertex correction in b) and self-energy corrections in c) and d).  The self-energy corrections are, for term c), $\Prop{n+1}\Sigma_{n+1}\Prop{n+1}\Ab\Prop{n}$, while for term d) we have $\Prop{n+1}\Ab\Prop{n}\Sigma_{n}\Prop{n}$. Here $\Sigma_{n}$ is the ultraviolet pole of the self-energy for an electron of momentum $p+nk$ in the Lorentz class. This can be directly evaluated using similar arguments to those used in Section 18.2.1 of~\cite{Schwartz:2013pla}. A more detailed discussion of the required integrals can be found from page 67 onwards  of~\cite{Pascual:1984zb}. Extracting the ultraviolet pole, we find
\begin{equation}\label{eq:sigma_n}
 	\Sigma_n=\big(i3m+\xi\Prop{n}^{-1}\big)\duv\,,
 \end{equation} 
 where, in dimensional regularisation with $D=4-2\varepsilon$ dimensions, 
\begin{equation}\label{eq:deltaUV}
  \duv=-\frac{e^2}{(4\pi)^2}\frac1{\varepsilon}\,.
\end{equation}
 
An ultraviolet, vertex correction is shown in Fig.~\ref{fig:in_plus_loop}~b). We denote this by $\Sigma_{\mathrm{in}}$, so that  Fig.~\ref{fig:in_plus_loop}~b) becomes equal to $\Prop{n+1}\Sigma_{\mathrm{in}}\Prop{n}$. We find  $\Sigma_{\mathrm{in}}$ by extracting the ultraviolet pole from the loop contribution 
 \begin{equation}
 	-ie^2 \!\intfsDuv \frac{\gamma^\mu (\ssl+\psl+(n+1)\ksl+m) \lasersl (\ssl+\psl+n \ksl+m)\gamma^\nu}{\big((s+p+(n+1)k)^2-m^2 +i\epsilon\big)  \big((s+p+nk)^2-m^2+i\epsilon \big)}D_{\mu\nu}(s)\,. 
 \end{equation}
Expanding the numerator in powers of the loop momenta,  the ultraviolet pole comes from the highest power. We thus obtain 
\begin{equation}
	\Sigma_{\mathrm{in}}= -\xi \Ab \duv\,.
\end{equation}
In a similar way, the loop correction to an emission into the background is found to be $\Sigma_{\mathrm{out}} = -\xi \hspace{.13pt}\Em \duv $.

Using the absorption identity (\ref{eq:AbsId}) allows us to write this vertex correction, in terms of sidebands, as
\begin{equation}
	\Sigma_{\mathrm{in}}=\xi\big(\In\Prop{n}^{-1}-\Prop{n+1}^{-1}\In\big)\duv
	=\In\Sigma_n-\Sigma_{n+1}\In\,.
\end{equation}
Note that the $\xi$-independent terms in the self-energies cancel in the last step.
The dual emission version of this is then
\begin{equation}
	\Sigma_{\mathrm{out}}=\Sigma_{n}\Out-\Out\Sigma_{n+1}\,.
\end{equation}
  
We thus see that, in the full Lorentz class of gauges, the absorption processes given in Fig.~\ref{fig:in_plus_loop} have the sideband expansion 
\begin{align}
\begin{split}
	\Prop{n+1}\Ab\Prop{n}+&\Prop{n+1}\Sigma_{\mathrm{in}}\Prop{n}+\Prop{n+1}\Ab\Prop{n}\Sigma_{n}\Prop{n}+\Prop{n+1}\Sigma_{n+1}\Prop{n+1}\Ab\Prop{n}\\
	&=\In\big(\Prop{n}+\Prop{n}\Sigma_n\Prop{n}\big)-
	\big(\Prop{n+1}+\Prop{n+1}\Sigma_{n+1}\Prop{n+1}\big)\In\,.
\end{split}
\end{align}
In this sideband description we see that each of the tree level propagators acquires the expected gauge dependent, one loop correction in the Lorentz class of gauges. The dual emission results are seen to have the same renormalisation structure. 

\begin{figure}[htb]
\[
\includegraphics{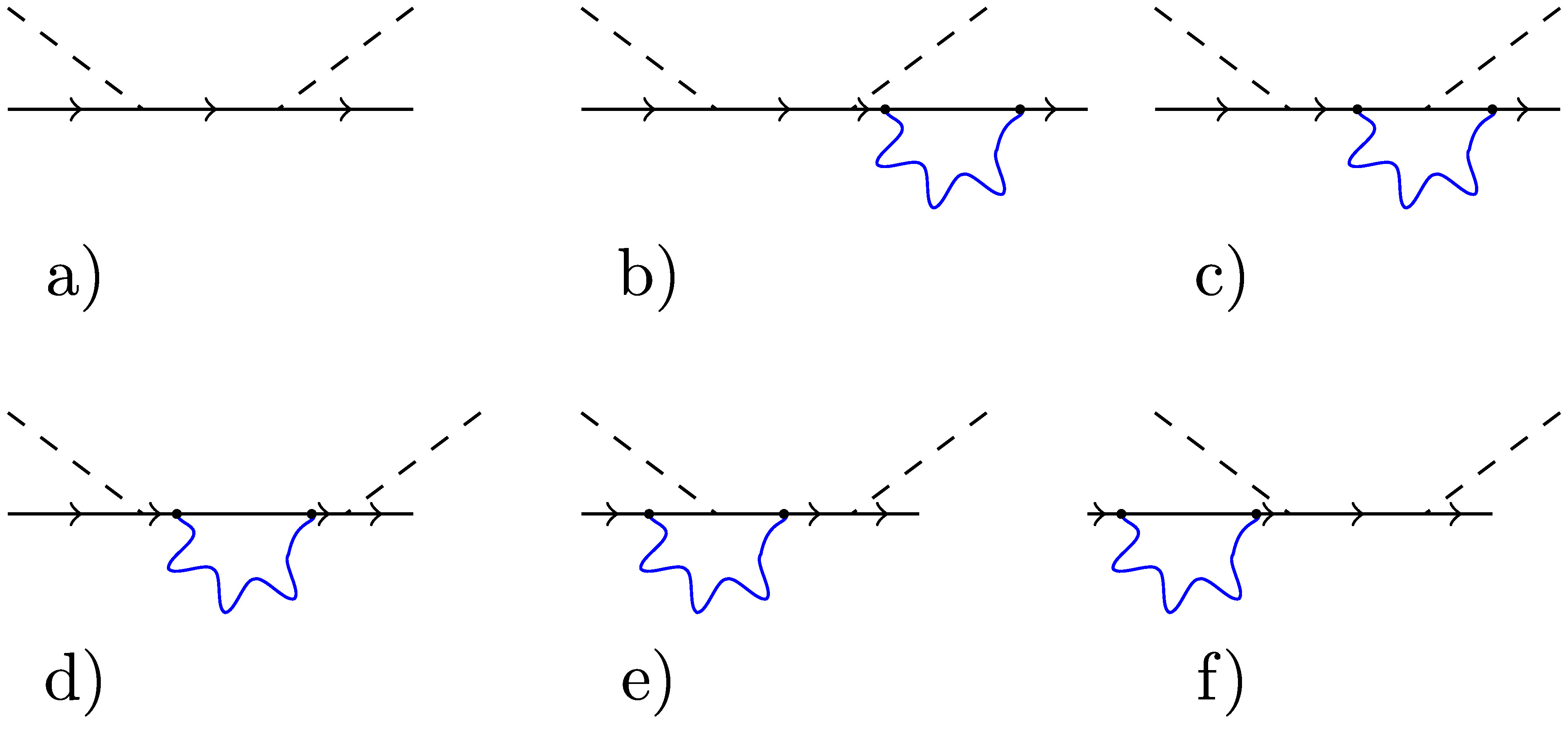}	
\] 
\caption{Absorption then emission at tree level and one-loop.}
	\label{fig:in_out_plus_loop}
\end{figure}

\begin{figure}[htb] 
\[
	\includegraphics{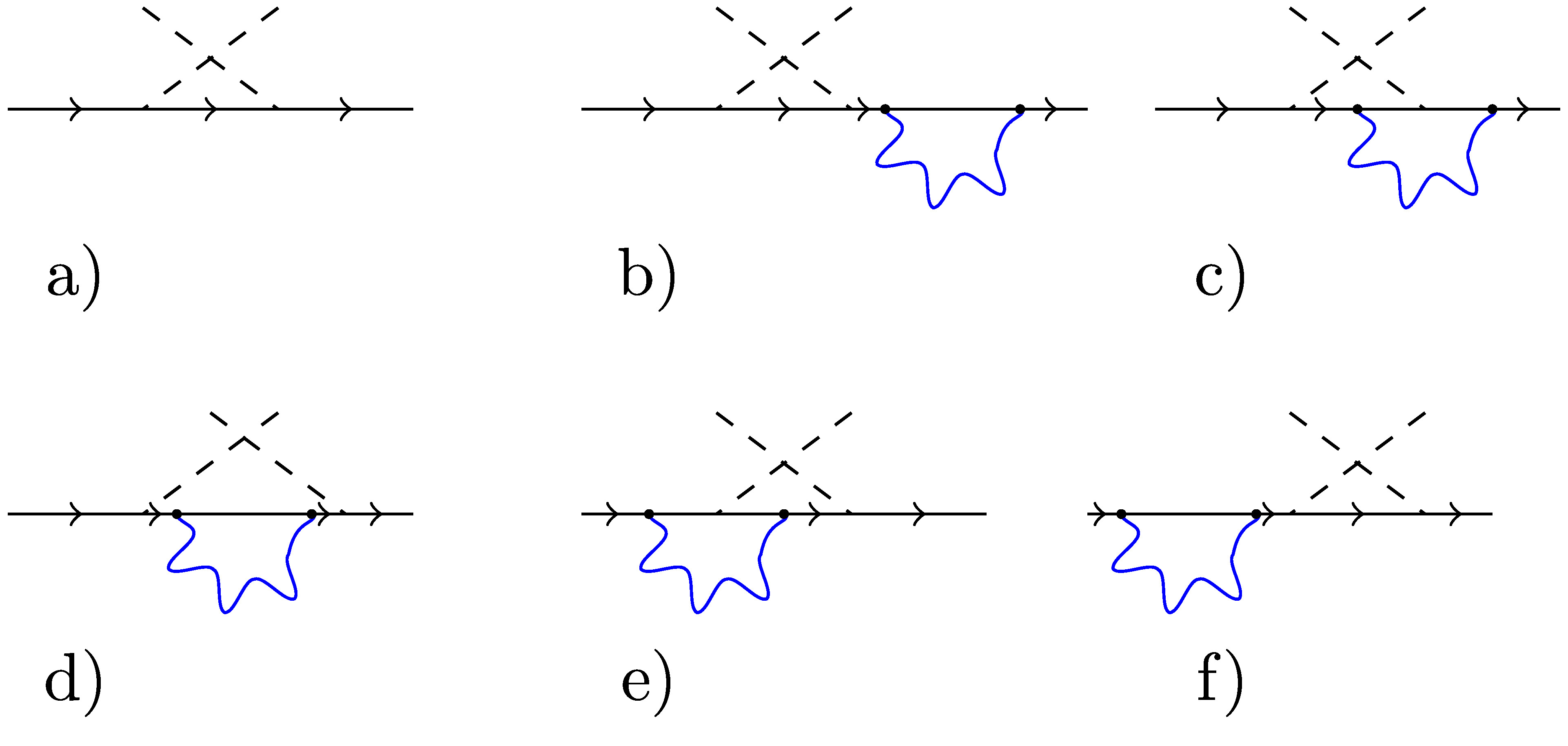}
\]
\caption{Emission then absorption at tree level and one-loop.}
	\label{fig:out_in_plus_loop}
\end{figure}

We now extend this analysis to the case of absorption and emission. This is important as such processes generate the laser induced mass shift. The relevant diagrams are given in Figs.~\ref{fig:in_out_plus_loop} and~\ref{fig:out_in_plus_loop}. Note that we have not included loops spanning more than one vertex as their contributions are ultraviolet finite. 
 
The tree level processes, Fig.~\ref{fig:in_out_plus_loop}a) and~Fig.~\ref{fig:out_in_plus_loop}a), generate the following sideband structures:
\begin{equation}\label{eq:central_sidebands}
\Prop{n}\Em\Prop{n+1}\Ab\Prop{n}+\Prop{n}\Ab\Prop{n-1}\Em\Prop{n}= \In\Prop{n-1}\Out-\Out\In\Prop{n}-\Prop{n}\Out\In-\Prop{n} i\mstarsl\Prop{n}+\Out\Prop{n+1}\In\,. 
\end{equation}
We see that they depend on three propagators,  $\Prop{n}$, $\Prop{n\pm1 }$, as well as a new  double pole, mass generating term
\begin{equation}\label{eq:vector_mass}
  \mstar_\mu:
  =-\frac{\laserm}{\pk}k_\mu\,, 
\end{equation}
which we note is independent of the polarisation. 
The rest of the diagrams contain the ultraviolet divergent one loop corrections to this.  The identity
\begin{equation}
  \Out\Sigma_{n+1}\In+\In\Sigma_{n-1}\Out=\Out\In\Sigma_{n}+\Sigma_{n}\Out\In-i\xi \mstarsl\duv\,,
\end{equation}
is key to the evaluation of these contributions. This result is the Lorentz class generalisation of equation (39) in \cite{Lavelle:2019lys}, and can be derived using the methods outlined in that paper. Note that every term in this expression contains an implicit or explicit  dependence on the  gauge parameter $\xi$.

Using these ultraviolet results,  the processes in Fig.~\ref{fig:in_out_plus_loop} and Fig.~\ref{fig:out_in_plus_loop}  can be shown to have the sideband expansion in the general Lorentz class of gauges given by
\begin{align}\label{eq:ren_inout}
\begin{split}
 &\In\Big(\Prop{n-1}+\Prop{n-1}\Sigma_{n-1}\Prop{n-1} \Big)\Out\\
 &\qquad-\Out\In\Big(\Prop{n}+\Prop{n}\Sigma_{n}\Prop{n} \Big) -\Big(\Prop{n}+\Prop{n}\Sigma_{n}\Prop{n} \Big)\Out\In\\
 &\qquad\qquad\qquad-\Big(\Prop{n}+\Prop{n}\Sigma_{n}\Prop{n}\Big)i\big(\mstarsl+\xi\sigmasl\big)\Big(\Prop{n}+\Prop{n}\Sigma_{n}\Prop{n}\Big)\\
&\qquad\qquad+\Out\Big(\Prop{n+1}+\Prop{n+1}\Sigma_{n+1}\Prop{n+1} \Big)\In\,,
\end{split}
\end{align} 
where
\begin{equation}\label{eq:massUV}
  \sigmasl=\frac{e^2}{(4\pi)^2}\frac1\varepsilon \mstarsl
  \,,
\end{equation}
is the one loop, laser induced mass correction previously found in the Feynman gauge,  see  \cite{Lavelle:2019lys}.

From the form of the self-energy (\ref{eq:sigma_n}), we see that in~(\ref{eq:ren_inout}) the wavefunction renormalisation in the individual sidebands is gauge dependent as in the vacuum theory. Similarly the usual double pole mass shift in each of the sidebands is gauge independent, again just as in vacuum. What is new and unexpected in~(\ref{eq:ren_inout}) is the explicit gauge dependence of the loop correction to the laser induced mass shift. 

Note that in the full class of gauges considered in this paper, the theory can be renormalised by introducing an additional counterterm for the laser induced mass shift. This was previously seen in Feynman gauge in \cite{Lavelle:2019lys}. 

It is interesting to note that in Landau gauge, $\xi=0$, this additional counterterm is not needed. The only loop corrections, and hence renormalisation, are those of the vacuum theory. This agrees with the results found by Brouder~\cite{Brouder:2002fz} who used a dressing approach~\cite{Bagan:1999jf}\cite{Bagan:1999jk} which effectively kept his calculations in Landau gauge.  
 
A possible reason why Landau gauge plays this special role is the fact that two gauge fixing conditions are being used in this theory: the Lorentz class in the loop calculation and the background gauge condition that $k{\cdot}\laser=0$. This last condition on the background corresponds to a Lorenz gauge choice for the single mode making up the background field.  In terms of the photon propagator, this corresponds to $\xi=0$, which is normally referred to as Landau gauge in perturbative calculations. This could account for the unique property of Landau gauge in the loop calculation. Note that one could equally interpret the background gauge condition as a light cone gauge fixing, which would suggest that if we had carried out the loop calculation in the light cone gauge, then again vacuum renormalisation might have sufficed. 
   
The results presented here should be contrasted with the gauge invariant photon polarisation. This has been calculated, see for example \cite{Becker:1974en}\cite{baier75a}\cite{Milstein:2006zz}\cite{Dinu:2013gaa}\cite{Meuren:2013oya} \cite{Gies:2016yaa}, in various backgrounds and at higher orders. Those calculations confirm that for these processes the vacuum subtraction suffices.

Finally we note that our analysis has focused on the ultraviolet structure of the theory and its renormalisation in this full class of gauges. Future work is needed to calculate the finite parts and the infrared structure of this theory.

\section*{Acknowledgements}
We wish to thank Tom Heinzl, Anton Ilderton and Ben King for discussions. 

\bigskip  
%\bibliographystyle{elsarticle-num}
%\bibliography{ref_database_2019}

\end{document}